\magnification 1200

%
%
\newdimen\FigSize       \FigSize=.9\hsize 
%
\newskip\abovefigskip   \newskip\belowfigskip
\gdef\epsfig#1;#2;{\par\vskip\abovefigskip\penalty -500
   {\everypar={}\epsfxsize=#1\noindent
    \centerline{\epsfbox{#2}}}%
    \vskip\belowfigskip}%
%
\newskip\figtitleskip
\gdef\tepsfig#1;#2;#3{\par\vskip\abovefigskip\penalty -500
   {\everypar={}\epsfxsize=#1\noindent
    \vbox
      {\centerline{\epsfbox{#2}}\vskip\figtitleskip
       \centerline{\figtitlefont#3}}}%
    \vskip\belowfigskip}%
%
\newcount\FigNr \global\FigNr=0
\gdef\nepsfig#1;#2;#3{\global\advance\FigNr by 1
   \tepsfig#1;#2;{Figure\space\the\FigNr.\space#3}}%
%
%
%
\gdef\ipsfig#1;#2;{
   \midinsert{\everypar={}\epsfxsize=#1\noindent
              \centerline{\epsfbox{#2}}}%
   \endinsert}%
%
\gdef\tipsfig#1;#2;#3{\midinsert
   {\everypar={}\epsfxsize=#1\noindent
    \vbox{\centerline{\epsfbox{#2}}%
          \vskip\figtitleskip
          \centerline{\figtitlefont#3}}}\endinsert}%
%
\gdef\nipsfig#1;#2;#3{\global\advance\FigNr by1%
  \tipsfig#1;#2;{Figure\space\the\FigNr.\space#3}}%
\newread\epsffilein    
\newif\ifepsffileok    
\newif\ifepsfbbfound   
\newif\ifepsfverbose   
\newdimen\epsfxsize    
\newdimen\epsfysize    
\newdimen\epsftsize    
\newdimen\epsfrsize    
\newdimen\epsftmp      
\newdimen\pspoints     
\pspoints=1bp          
\epsfxsize=0pt         
\epsfysize=0pt         
\def\epsfbox#1{\global\def\epsfllx{72}\global\def\epsflly{72}%
   \global\def\epsfurx{540}\global\def\epsfury{720}%
   \def\lbracket{[}\def\testit{#1}\ifx\testit\lbracket
   \let\next=\epsfgetlitbb\else\let\next=\epsfnormal\fi\next{#1}}%
\def\epsfgetlitbb#1#2 #3 #4 #5]#6{\epsfgrab #2 #3 #4 #5 .\\%
   \epsfsetgraph{#6}}%
\def\epsfnormal#1{\epsfgetbb{#1}\epsfsetgraph{#1}}%
\def\epsfgetbb#1{%
%
%
\openin\epsffilein=#1
\ifeof\epsffilein\errmessage{I couldn't open #1, will ignore it}\else
%
%
   {\epsffileoktrue \chardef\other=12
    \def\do##1{\catcode`##1=\other}\dospecials \catcode`\ =10
    \loop
       \read\epsffilein to \epsffileline
       \ifeof\epsffilein\epsffileokfalse\else
%
%
          \expandafter\epsfaux\epsffileline:. \\%
       \fi
   \ifepsffileok\repeat
   \ifepsfbbfound\else
    \ifepsfverbose\message{No bounding box comment in #1; 
   using defaults}\fi\fi
   }\closein\epsffilein\fi}%
%
%
\def\epsfsetgraph#1{%
   \epsfrsize=\epsfury\pspoints
   \advance\epsfrsize by-\epsflly\pspoints
   \epsftsize=\epsfurx\pspoints
   \advance\epsftsize by-\epsfllx\pspoints
%
%
   \epsfxsize\epsfsize\epsftsize\epsfrsize
   \ifnum\epsfxsize=0 \ifnum\epsfysize=0
      \epsfxsize=\epsftsize \epsfysize=\epsfrsize
%
%
     \else\epsftmp=\epsftsize \divide\epsftmp\epsfrsize
       \epsfxsize=\epsfysize \multiply\epsfxsize\epsftmp
       \multiply\epsftmp\epsfrsize \advance\epsftsize-\epsftmp
       \epsftmp=\epsfysize
       \loop \advance\epsftsize\epsftsize \divide\epsftmp 2
       \ifnum\epsftmp>0
          \ifnum\epsftsize<\epsfrsize\else
             \advance\epsftsize-\epsfrsize \advance\epsfxsize\epsftmp 
     \fi
       \repeat
     \fi
   \else\epsftmp=\epsfrsize \divide\epsftmp\epsftsize
     \epsfysize=\epsfxsize \multiply\epsfysize\epsftmp   
     \multiply\epsftmp\epsftsize \advance\epsfrsize-\epsftmp
     \epsftmp=\epsfxsize
     \loop \advance\epsfrsize\epsfrsize \divide\epsftmp 2
     \ifnum\epsftmp>0
        \ifnum\epsfrsize<\epsftsize\else
           \advance\epsfrsize-\epsftsize \advance\epsfysize\epsftmp \fi
     \repeat     
   \fi
%
%
   \ifepsfverbose\message{#1: width=\the\epsfxsize, 
   height=\the\epsfysize}\fi
   \epsftmp=10\epsfxsize \divide\epsftmp\pspoints
   \vbox to\epsfysize{\vfil\hbox to\epsfxsize{%
      \includegraphics{#1}%
      \hfil}}%
\epsfxsize=0pt\epsfysize=0pt}%
%
%
{\catcode`\%=12 
 \global\let\epsfpercent=
%
%
\long\def\epsfaux#1#2:#3\\{\ifx#1\epsfpercent
   \def\testit{#2}\ifx\testit\epsfbblit
      \epsfgrab #3 . . . \\%
      \epsffileokfalse
      \global\epsfbbfoundtrue
   \fi\else\ifx#1\par\else\epsffileokfalse\fi\fi}%
%
%
\def\epsfgrab #1 #2 #3 #4 #5\\{%
   \global\def\epsfllx{#1}\ifx\epsfllx\empty
      \epsfgrab #2 #3 #4 #5 .\\\else
   \global\def\epsflly{#2}%
   \global\def\epsfurx{#3}\global\def\epsfury{#4}\fi}%
%
%
\def\epsfsize#1#2{\epsfxsize}%
%
%

\epsfverbosetrue                        
\abovefigskip=\baselineskip             
\belowfigskip=\baselineskip             
\global\let\figtitlefont\bf             
\global\figtitleskip=.5\baselineskip    

\font\tenmsb=msbm10   
\font\sevenmsb=msbm7
\font\fivemsb=msbm5
\newfam\msbfam
\textfont\msbfam=\tenmsb
\scriptfont\msbfam=\sevenmsb
\scriptscriptfont\msbfam=\fivemsb

\let\nd\noindent 

\def\natural{{\rm I\kern-.18em N}}

\def\chix{{\raise.5ex\hbox{$\chi$}}}
\def\chixa{{\chix\lower.2em\hbox{$_A$}}}

\def\real{{\rm I\kern-.2em R}}
\def\integer{{\rm Z\kern-.32em Z}}
\def\complex{\kern.1em{\raise.47ex\hbox{
            $\scriptscriptstyle |$}}\kern-.40em{\rm C}}
\def\vs#1 {\vskip#1truein}
\def\hs#1 {\hskip#1truein}

\def\Month{\ifcase\number\month \relax\or January \or February \or
  March \or April \or May \or June \or July \or August \or September
  \or October \or November \or December \else \relax\fi }
\def\date{\Month \the\day, \the\year}

  \hsize=6.5truein        \hoffset=0truein 
  \vsize=8.8truein      
  \pageno=1     \baselineskip=12pt
  \parskip=0 pt         \parindent=20pt
  \overfullrule=0pt     \lineskip=0pt   \lineskiplimit=0pt
  \hbadness=10000 \vbadness=10000 
\pageno=0

\footline{\ifnum\pageno=0\hss\else\hss\tenrm\folio\hss\fi}
\hbox{}
\vskip 1truein\centerline{{\bf A Revolutionary Material}}
\vskip .1truein\centerline{by}
\vskip .1truein
\centerline{{Charles Radin
\footnote{*}{Research supported in part by NSF Grant DMS-1208941}}}
\vskip .2truein\centerline{Department of Mathematics} 
\vskip 0truein\centerline{University of  Texas} 
\vskip 0truein\centerline{Austin, TX\ \ 78712}

\vs1 
\centerline{{\bf Abstract}}
\vs.1 \nd
This is an expository introduction, for a general mathematics 
audience, to the modeling of the fluid/solid phase transition and in
particular to complications created by the discovery of quasicrystals.
One goal is to elucidate certain features of the modeling which are
ripe for mathematical investigation.


\vfill\eject
The 2011 Nobel prize for chemistry was awarded to Dan Shechtman for
the discovery of quasicrystals, an exotic class of materials. The
discovery was published in 1984 and was quickly treated as
revolutionary, with front page headlines in newspapers.

While the award was for chemistry, the revolution was more broadly
based within the interdisciplinary subject of materials science. This
can be described easily and we will begin with a sketch of this. The
multifaceted implications for mathematics are more complicated and we
will try to elucidate them afterwards.

The basic fact is that quasicrystals are equilibrium solids which are
not crystalline. Not only is their pattern of atoms not crystalline,
the pattern has a fascinating hierarchical structure. However we
emphasize that the hierarchical pattern is not essential to the
revolutionary significance of quasicrystals to materials science.

It had been understood for many years, following the development of
X-ray diffraction, that common inorganic solids (for instance all
solids composed of only one chemical element) are crystalline, and
great practical success followed from incorporating this in their
modeling, essentially by analyzing various perturbations of a
crystalline atomic configuration. This is evident from standard
textbooks on solid state physics from the 1970's. The startling fact
uncovered by the discovery of quasicrystals was the existence of a
previously unknown class of inorganic solids, of unknown diversity,
for which a fundamentally different approach would be needed,
specifically without the help of an underlying crystalline
structure. That was the revolution in materials science.

As for the implications for mathematics, one path quickly developed
from the hierarchical atomic patterns, which played a central role in the
theory of  Levine and Steinhardt based on aperiodic tilings such
as the Penrose `kites and darts' (see Figure 1). 
\vs.1
\epsfig .47\hsize; 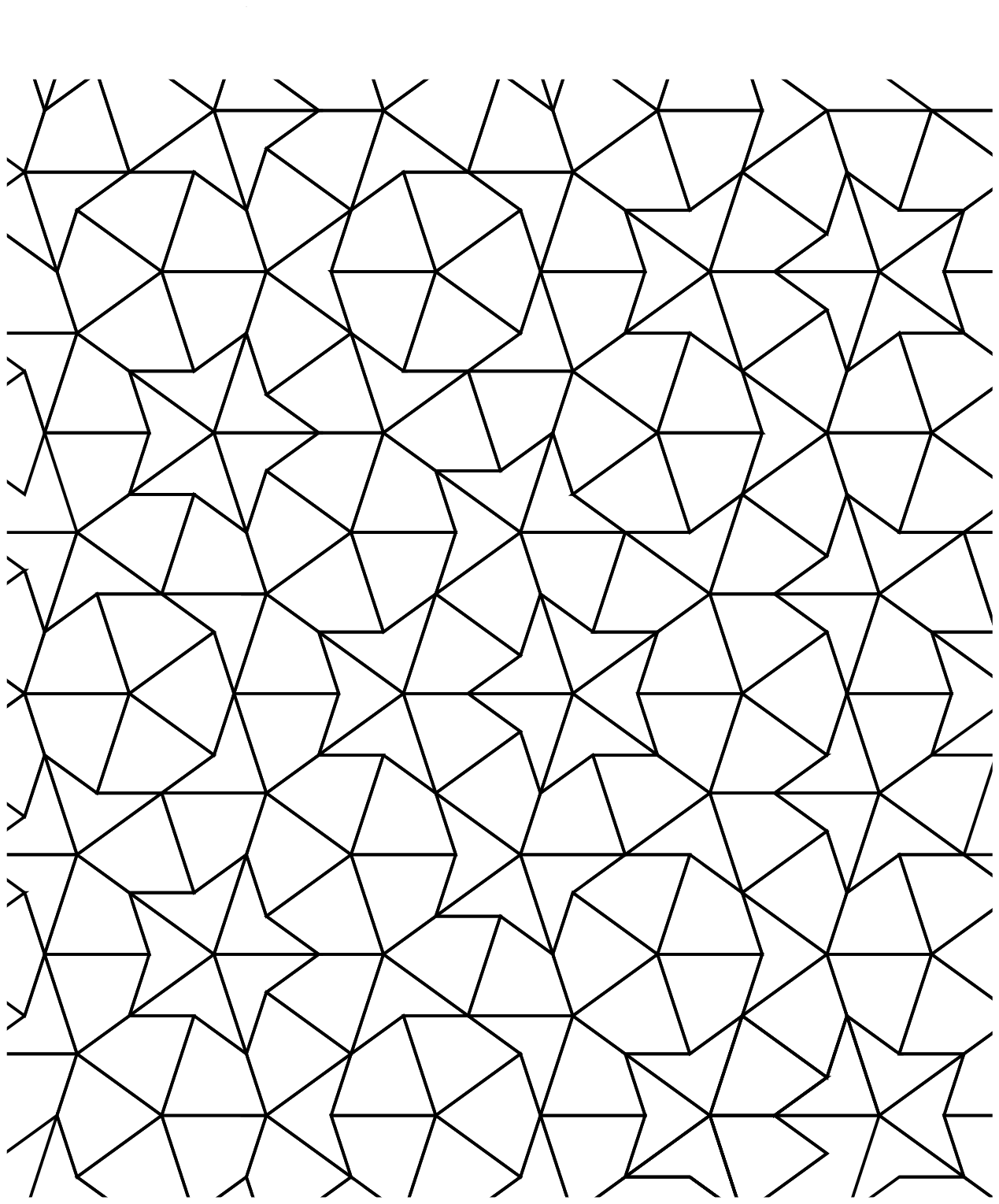 ;
\vs0
\centerline{Figure 1. A Penrose kite and dart tiling of the plane}
\vs.1

This led to
interesting mathematics; the developments with which I am familiar
were in the ergodic theory of aperiodic tilings, and the theory
of density in hyperbolic spaces, and there are undoubtedly many
more developments inspired by the hierarchical structure.

That was the first, early, line of development in mathematics coming
from quasicrystals. But there is also significant mathematics
intimately related to the revolutionary character of quasicrystals,
the basic fact that quasicrystals are noncrystalline
solids. Clarifying this mathematical connection is the goal for
the rest of this article. This will require some review of the nature
of solids and their modeling using equilibrium statistical mechanics,
which we will motivate by focusing on a certain phenomenon.

To understand a material scientifically one typically studies the experimental
response to a disturbance; one kicks it and examines the
reaction. Electrical conduction concerns the response to an applied
voltage; sound propagation concerns the response to a (rapidly
varying) applied pressure; elastic stress coefficients model the
response to applied mechanical forces, and so on. Let us explore mechanical
forces in more depth, through the following specific problem.

It is natural that if you tried to stand on an ocean surface you would
sink because the water molecules would move out of the way of the
force applied by your feet. Why then can you stand on a glacier? It
turns out that the more one analyzes these two contrasting material
responses, of water and of ice, the more intriguing the question
becomes, and we will use this to clarify the
significance within mathematics of the discovery of quasicrystals. So
we will keep in mind the problem:
\vs.1
\item{(1)}\hfill Why can you stand on ice but not water? \hfill \hbox{}
\vs.1

We begin with a certain classification
of applied mechanical forces, or stresses. Suppose we have one
balloon filled with water and another filled by a single block of
ice. The possible stresses we could apply to the balloons are commonly
classified into `pressure', which tends to change the volume but not
the shape of the balloon, and `shear stress', which tends to change the
shape but not the volume. Shape is quantified by angles denoting
`strain'; see Figure 2. 
\vs.2
\epsfig .7\hsize; 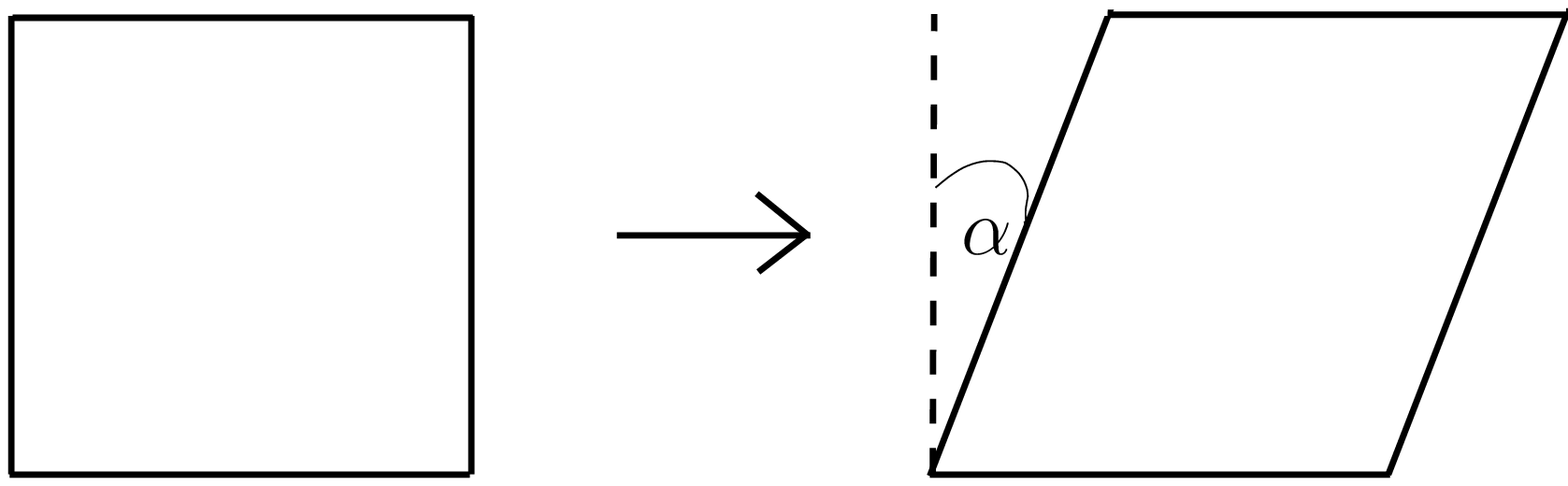;
\vs0 \nd
\centerline{Figure 2. A strain angle $\alpha$}
\vs.1 \nd
We can view the force and the associated
geometric change as responses to one another; applying a force
yields a change in geometry of the bulk material, and effecting a
change in the geometry is resisted by a corresponding force from the
material. The response is generally a nonlinear function of its
cause, but the coefficient of the linear approximation
is useful. The linear coefficient of response forces to
changes in geometry are called moduli: (elastic) bulk modulus for
pressure, and (elastic) shear modulus for shear.

Getting back to our question (1) and the need to distinguish the
response of ice from that of water, we choose to concentrate on shear,
in particular the shear moduli of ice and water. Pressure would be
much simpler to analyze but of little value since water is an
incompressible fluid, with almost the same bulk modulus as ice. But
water deforms rather than support any (static) shear at all, while ice
is hard to deform, so the shear modulus of water is zero while it is
large for ice. So to answer (1) we will try to understand through models 
why the shear
moduli of water and ice are so different. And perhaps we can also
reverse focus and ask whether this difference is the key to the
fundamental difference between water and ice.

We will explore our problem now in more depth, beginning with the
thermodynamic model of matter as an intermediate step towards the
statistical mechanics model. For convenience we will restrict our
modeling to `simple' materials, which are
(macroscopically) homogeneous, isotropic, uncharged and of only one 
chemical species, and that are not acted on by magnetic, electric or
gravitational forces. The model will thus be restricted to questions
of internal energy, such as the transfer of energy between two systems
in contact, and the interaction of these with mechanical operations on
the systems. A typical application might concern the energy of a gas
in the chamber of a piston, the whole bathed in a fluid at fixed
temperature, when the chamber of the piston is expanded.

The formalism of thermodynamics makes use of a quantity called the
entropy density of the system. Experiment demonstrates that a simple
system can be put into thermal equilibrium, where it has a range of
well-defined equilibrium states, parameterized in several equivalent
ways but for instance by the two quantities of energy density $e$ and
mass density $m$, so that all thermodynamic quantities, including the
entropy density and the various mechanical properties, have unique
values for given ($e,m)$. Furthermore it is found that all
thermodynamic properties are computable from the entropy density
function, $s(e,m)$. For instance $\displaystyle {\partial s/\partial
e}$ is inversely proportional to the temperature. A transform of
$s(e,m)$, called the Gibbs free energy density $g(P,T)$, basically a
Legendre transform of $s(e,m)$, can play a role alternative to
$s(e,m)$ but with variables $P,T$, the pressure and temperature,
respectively.
\vs.1
\item{(2)} All thermodynamic properties are uniquely determined by the
entropy density $s(e,m)$, or alternatively by the Gibbs free energy
density $g(P,T)$.
\vs.1 \nd
We next show how useful this can be in the modeling of mechanical properties.

The experimental states of bulk matter in thermal equilibrium can be organized into
`phases'. A phase is the set of states
in an open subset of the parameter space $\{(P,T)\}$ which
is maximal with respect to the property that within that subset all
thermodynamic properties are analytic. From (2) it suffices to
require this of just $g(P,T)$. So the boundary of a phase consists of
singularities of $g(P,T)$.

The simplest phase of any material is the (isotropic) fluid phase, the
phase which contains all $(P,T)$ with $P$ sufficiently low and $T$
sufficiently high. The complement of the fluid phase, for any
material, contains those $(P,T)$ with $P$ sufficiently high and $T$
sufficiently low, and contains one or more distinct `solid' phases,
typically with distinct crystal structure. See Figure 3, which
includes the curves of singularities of $g(P,T)$, bounding the fluid
and solid phases. We note that a phase can bound itself; in fact the
part of the boundary of the fluid phase at which that phase bounds
itself is where the gas and liquid forms of the fluid phase
coexist; see Figure 3.

\vs.1
\epsfig .4\hsize; 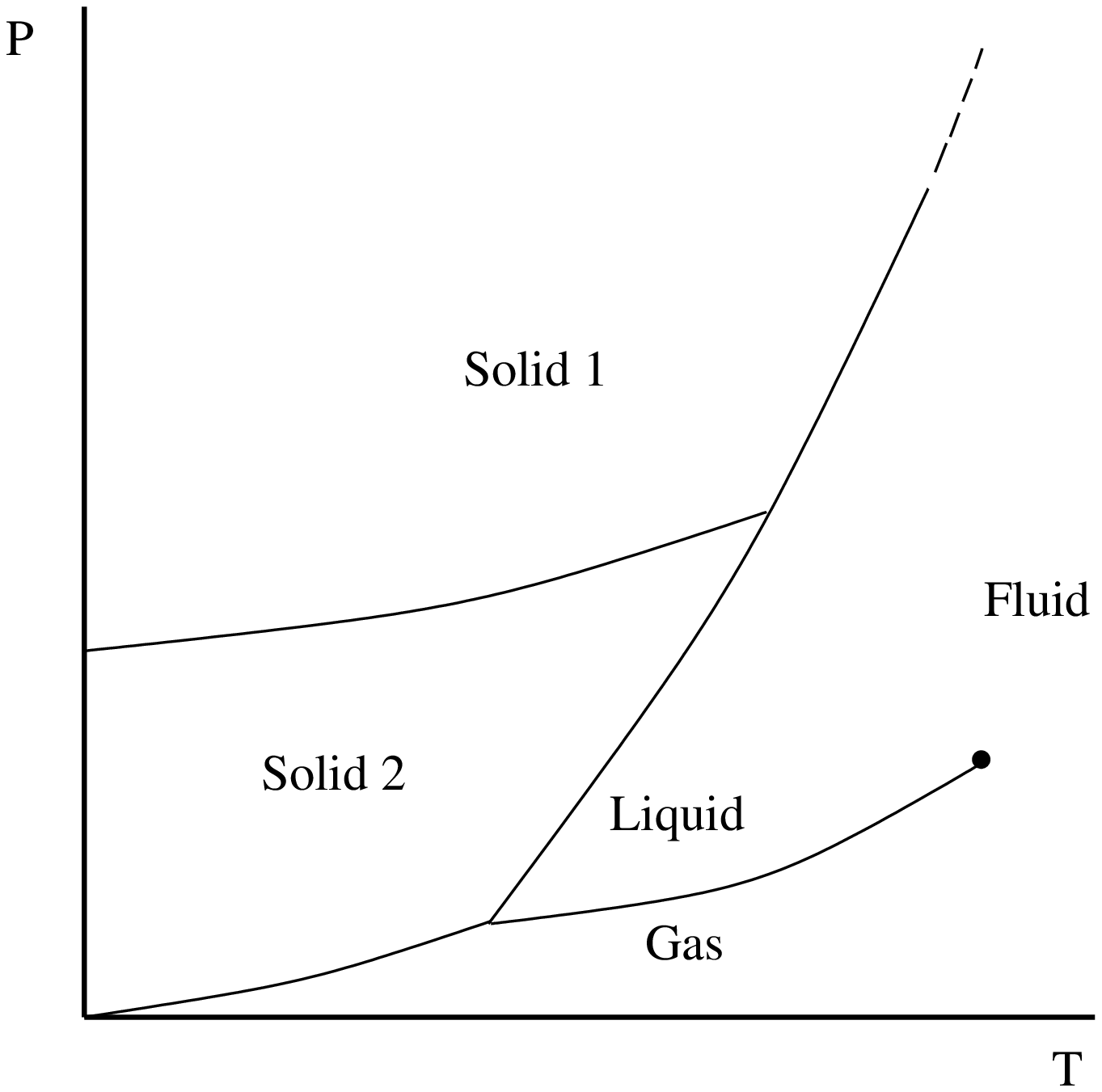;
\vs0 \nd
\centerline{Figure 3. A schematic experimental parameter space}
\vs.1

Using the language of phases our problem is to understand why the
shear moduli of the fluid and solid phases of water are so
different. We sketched the thermodynamic analysis of simple matter,
but this formalism does not address the causes of the phenomena it
describes; traditionally one goes to a deeper level of analysis,
equilibrium statistical mechanics, for such understanding.

In brief, statistical mechanics tries to show how the thermodynamic
properties of a material follow from the {\it interaction} of
constituent particles, which we will call molecules; note that water and ice are
both made from $\displaystyle H_2O$ molecules, with the same
interaction. The pressure of a
gas, which has a natural meaning in terms of the mechanical properties
of the bulk material, can also be understood in terms of momentum
transfer between the molecules and between the molecules and the environment
of the gas. The internal energy of the system can be understood in
terms of the mechanical notions of potential and kinetic
energies. However the biggest step in the development of statistical
mechanics, due to Boltzmann, was the model of entropy density as
$\displaystyle (1/V) \ln[\Gamma_V(e,m)]$, where $V$ is the volume of
the material in space and $\Gamma_V(e,m)$ is the (high dimensional)
volume of the set of all joint states $c$ of the molecules which have
total (kinetic plus potential) energy density $e(c)$ and mass density
$m(c)$. The advantage of such modeling is that if we can compute the
potential and kinetic energies between the molecules (even in a model with
unrealistic interactions) we could in principle compute the entropy
density $s(e,m)$ (or Gibbs free energy density $g(P,T)$) from which
all thermodynamic information would follow. This would provide a
deeper understanding of the thermodynamic properties; the
all-important function $s(e,m)$, or $g(P,T)$, would follow from the
interactions of the component molecules.

One feature of statistical mechanics which was omitted above, but
which we need, concerns the phases. Since we are trying to understand
the fundamental difference between these phases, it is necessary that
we have access to them in our modeling. Now it has been known for many
years that if we accept the conventional meaning of phase boundaries
as singularities of $s(e,m)$ or $g(P,T)$ we must take the limit of
system size to infinity in the models. So to model the entropy density
we use:
$$s(e,m) = \lim_{V\to \infty} {1\over
V} \ln[\Gamma_V(e,m)]. \leqno{(3)}  $$

The above was a superficial introduction to the standard modeling of
bulk materials in thermal equilibrium, including the notions of fluid
and solid phases, but even though superficial we can see that this
modeling is insufficient to deal with our problem of the rigidity of
solids. The difficulty is that the above theory does not address the
response of a material to an applied shear strain! In fact it has been
proven that the above entropy density $s(e,m)$ and Gibbs free energy
density $g(P,T)$ are independent of the shape of the material, and
indeed we can look up material properties without specifying the shape
of our material sample. And if $g(P,T)$ is independent of shape how
can we compute from it a response to changing the shape? This is our
problem: for a macroscopic system of interacting molecules in thermal
equilibrium, how do we model the response to shear strain, and in
particular show that it is high for state parameters $(P,T)$
corresponding to the solid phase but identically zero for $(P,T)$
corresponding to the fluid phase?

We will sketch two approaches to this. The first, by Aristoff and
myself, gets around the above difficulty by three steps. One is the
observation that the response to strain is in fact computable using the
statistical mechanics of a finite system, before taking the limit in
system size. The next idea is a bit technical, namely to use a
response of a simpler nature than the reaction force of the system,
namely the volume or mass density of the system. That is, one changes the
shape and measures whether or not the volume changes.  This is mildly
counterintuitive since shear is not supposed to change volume, but
indeed it can, and this is actually a well-known phenomenon of sand, called
dilatancy. More formally it is reasonable if we note that because of
the singularities of the free energy all along the boundary separating
phases $g(P,T)$ is intuitively a completely different function in
different phases so we might well expect every thermodynamic quantity,
including the mass density, to be singular as $(P,T)$ crosses the boundary
between phases. The last idea is to look for a difference as we
interchange limits, namely the infinite size limit and the limit of
infinitely small strain implicit in the derivative of mass density with
respect to strain. More specifically, it is not hard to write a
formula, for a system of finite fixed size, for the linear response,
in other words the derivative, of the (average) mass density with
respect to shear strain. Then we can take the limit in the size of the
system. To repeat: we take the limit of vanishingly small shear {\it
before} we take the limit of infinite size. As noted above we know
that taking the limits in the other order cannot work because as we
take the infinite size limit the free energy loses its dependence on
the shape of the system; but does interchanging the limits help?  The
quantity we end up with, the volume limit of the derivative, is no
longer the linear coefficient in an expansion of a response, since
there is no response by the infinite system. But it still might be
meaningful. We focused on the response because we thought it might
distinguish water from ice, and the quantity we end up with may not be
easy to interpret as a response but it still could play a useful role
in distinguishing fluid from solid in models. Does it do this, and if it
does what does it represent physically?

I said we can write a formula for the response, but it is a
complicated integral, with parameters $P$, $T$ and $V$, in high
dimension and I did not say we could compute it analytically, or even
get useful qualitative information from it. The only evidence there is
for the above theory comes from simulation in a standard model called
`hard disks'. For that model one can let a computer (actually {\it
many} computers) apply Monte Carlo techniques to simulate the desired
equilibrium quantities, and the result is that the linear response of
mass density with respect to shear strain jumps from identically zero to
nonzero precisely (within error!) as the thermodynamic parameter
values cross the phase transition boundary between fluid and solid. So
it seems to work precisely as desired in an important model, though
this still leaves open its physical interpretation.

We now note a somewhat different approach to our basic problem (1),
distinguishing water from ice, by Sausset, Biroli and Kurchan,
which uses the response to a time dependent shear, a shear with
constant {\it strain rate}, a standard quantity when analyzing fluids. The
linear response of a fluid to a constant shear strain rate, namely the
(linear coefficient in the) response force to that deformation rate,
is called viscosity. The authors analyze the viscosity of crystals
using various traditional physics approximations and conclude that the
difference between a fluid and a solid is that within a solid the
viscosity diverges in the limit of zero shear strain, while within a
fluid the viscosity vanishes in the limit of zero shear strain. This
offers a different intuitive picture of the essential difference
between ice and water from that discussed first.

We cannot easily sketch the argument using viscosity because it
concentrates on time dependence which is difficult to mesh with a
well-defined notion of phase, which in modeling requires time
independent equilibrium systems of infinite size. But we have included
the approach specifically to bring up the important issue of time
dependence, both in the physical material and in models of
it. Throughout our discussion we have emphasized that quasicrystals
are materials in thermal equilibrium, meaning they have the stability
property that if perturbed by `annealing', the details of which are
unimportant here, the system would return to its original state as
measured by all thermodynamic properties. It is easy to prepare simple
dilute gases in thermal equilibrium; all that is needed is to provide
a steady environment of given pressure and temperature and the system
will naturally and quickly approach equilibrium. This is much harder
to do with solids, and in practice many solids would change their
state if annealed. Indeed it is common to purposely prepare solids out
of equilibrium in order to obtain desirable features; permanent
magnets are examples, as are (structural) glasses such as window
glass. From X-ray diffraction we know that the atomic positions in
window glass are indistinguishable from that of the material in a
liquid state at some $(P^\prime,T^\prime)$ corresponding to its
manufacture process, rather than being crystalline as it would be in
the true equilibrium state of the material at the $(P,T)$ of room
pressure and temperature. And yet of course window glass is quite
rigid. So a system which technically is just a very sluggish
(`viscous') fluid, not in thermal equilibrium, can behave like an
equilbrium solid. This makes (nonequilibrium) glasses notoriously
difficult to model. When we model a quasicrystal we can use the fact
that the material really is in thermal equilibrium, in effect that an
infinite time limit has been taken, and clearly in our modeling we
must analyze the proper order of taking that limit and the other
limits of interest, namely the (technically challenging) infinite size
limit and the limit of zero strain. The proper simultaneous handling
of these three limits is highly nontrivial and such modeling issues
cry out for the attention of serious applied mathematics. 
The traditional role for mathematics in open physics problems, for
instance concerning phase transitions, is to give proofs in standard,
simplified models. However because of the sophisticated technical
issues involved this problem seems to call for a different sort of
role for mathematics, namely in helping to determine the correct intuitive
understanding of the phenomenon at hand, the difference between fluids 
and solids in thermal equilibrium.

In summary, there is a fundamental open problem in condensed matter
physics to understand the essential difference between water and
ice. In physics language we are looking for the right `order
parameter' to distinguish the fluid and solid phases of matter in
thermal equilibrium. It is perhaps surprising that no one has ever
found an order parameter with which it could actually be proven, in
some simple but convincing model, that a molecular system has a sharp
transition between fluid and solid phases, so we could say that the
shear modulus (or alternatively the derivative of density with respect
to strain, or the viscosity, both sketched above) might play that
role. Many of the older attempts to find such an order parameter
focused on the difference in symmetry, the complete Euclidean symmetry
of the fluid versus the crystalline symmetry of the solid. The
existence of quasicrystals has affected this basic problem by showing
that crystalline symmetry, and by extension perhaps symmetry itself,
may not be relevant to an understanding of the fundamental difference
between fluid and solid phases, and this fact motivated the attempts
sketched above. The present physical theory of the wide range of phase
transitions of materials developed in part by motivating significant
progress in combinatorics and probability. Finally coming to grips
with this most fundamental of phase transitions, the fluid/solid
transition, may well require something different, a close
collaboration of mathematics and physics in the basic modeling, and
this is just one natural fallout of the quasicrystal revolution.
\vfill \eject
\centerline{\bf Some References for the Physics}
\vs.15 \nd
These are the initial reports on the experimental discovery, 
and theory, of quasicrystals.
\vs.1
\item{} D. Shechtman, I. Blech, D. Gratias and J.W. Cahn,
Metallic phase with long-ranged orientational order and no
tranlational symmetry,
{\it Phys. Rev. Lett.} {\bf 53} (1984) 1951-1953.
\vs.1
\item{} D. Levine and P.J. Steinhardt, 
{Quasicrystals: a new class of ordered structures},
{\it Phys. Rev. Lett.} {\bf 53} (1984) 2477-2480.
\vs.15 \nd
Here are standard references to thermodynamics and statistical mechanics.
\vs.1
\item{} H.B. Callen, {\it Thermodynamics} (John Wiley, New York, 1960).
\vs.1
\item{} S.-K. Ma, {\it Statistical Mechanics} (World Scientific,
Singapore, 1985).
\vs.15 \nd
Chapter 2 of the following contains an intriguing analysis of the role of `rigidity'
in understanding phase transitions.
\vs.1
\item{} P.W. Anderson, {\it Basic Notions of Condensed Matter Physics} 
(Benjamin/Cummings, Menlo Park, 1984).
\vs.15 \nd
These are the two papers, proposing theories of shear, which are
discussed in the article.
\vs.1
\item{} D. Aristoff and C. Radin, Rigidity in solids,
{\it J. Stat. Phys.} {\bf 144} (2011) 1247-1255.
\vs.1
\item{} F. Sausset, G. Biroli and J. Kurchan, Do Solids Flow?,
{\it J. Stat. Phys.} {\bf 140} (2010) 718-727.
\vs.15
\centerline{\bf Some References for the Mathematical Formalism}
\vs.15 \nd
These are {the} standard references on the mathematical control
of infinite size limits in statistical mechanics.
\vs.1
\item{}D. Ruelle, {\it Statistical Mechanics; Rigorous Results} (Benjamin,
New York, 1969).
\vs.1
\item{} D. Ruelle, {\it Thermodynamic Formalism} (Addison-Wesley, New
York, 1978). 
\vs.15 \nd
Following are two useful expository articles on mathematical issues
relevent to the article.
\vs.1
\item{}O.E. Lanford, Entropy and equilibrium states in classical statistical mechanics,
in {\it Statistical Mechanics and Mathematical Problems,
Battelle Seattle 1971 Rencontres, volume 20 of Springer Lecture Notes
in Physics}, ed. A. Lenard (Springer-Verlag, 
Berlin/Heidelberg/New York, 1973), pp. 1-113.
\vs.1
\item{}Robert B. Griffiths, Rigorous Results and Theorems, in
{\it Phase Transitions and Critical Phenomena, vol.\ 1},
ed. C. Domb and M.S. Green (Academic Press, New York, 1972), pp. 7-109.


\vfill \end